\documentclass[aip,graphicx]{revtex4-2}

\usepackage{amssymb}
\usepackage{mathrsfs}
\usepackage{stmaryrd}
\usepackage{subfigure}

\usepackage{multirow}
\usepackage{graphicx}
\usepackage{color}

\usepackage{enumerate}
\usepackage{fancyhdr}
\newcommand{\mb}{\mathbf}

\newcommand{\mr}{\mathrm}

\begin{document}
	
	\title{Dimensional Analysis Theory and Molecular Dynamics Simulation of Polypropylene Melt Flow during Injection Molding Process}
	
	\author{Jinrong Zhang}
	\affiliation{Department of Physics, Beijing Normal University, Beijing 100875, China}
	
	\author{Dadong Yan}
	\email{yandd@bnu.edu.cn}
	\affiliation{Department of Physics, Beijing Normal University, Beijing 100875, China}
	
	\author{Li Peng}
	\affiliation{National-certified Enterprise Technology Center, Kingfa Science and Technology Co., LTD., Guangzhou 510663, China}
	
	\author{Xianbo Huang}
	\affiliation{National-certified Enterprise Technology Center, Kingfa Science and Technology Co., LTD., Guangzhou 510663, China}
	
	\bigskip
	
	\begin{abstract}
	
	Flow marks are common surface defects that occur in injection-molded products. Their formation may be related to the flow process of the melt in the mold.\cite{hirano2007morphological,hirano2007striped,maeda2021tiger,maeda2021unusual,maeda2007flow} Through dimensional analysis, we have discovered that the geometric shape of the flow field is controlled by specific dimensionless quantities. These quantities can be summarized as follows: geometric dimensionless quantities related to the shape of the mold, material dimensionless quantities related to the melt and mold materials, and physical dimensionless quantities related to the flow. When the geometric shape of the mold changes proportionally, with the melt and mold material fixed, and the initial temperature of the melt and mold fixed, the geometric shape of the flow field will be solely controlled by the Weissenberg number $Wi$. If $Wi$ is kept constant, changing the injection speed, changing the relaxation time of the polypropylene melt, or scaling the mold will result in similar geometric shapes of the flow field. If the size of the mold is not changed, the geometric shape of the flow field will be the same. Since the dimensionless equation represents a similar system of all sizes, we verified the above conclusion through molecular dynamics simulations at a smaller scale. After further improvement of the micro simulation system, there is a possibility of visualizing the formation process of flow marks. This would greatly aid in the advancement of theory and the elimination of flow marks in production and experiments. This work also illustrates that the methodology of dimensional analysis plus molecular dynamics simulation may be applied to a wider range of other systems, scaling down large systems and thus significantly reducing their computational effort.

	\end{abstract}
	
	\maketitle
	
	\section{Introduction}

	Polypropylene (PP) mixtures made from melting and mixing ethylene propylene block copolymer (EPBC), ethylene propylene rubber (EPR), and ethylene propylene rubber (EBR), or inorganic fillers such as talc powder, are widely used in various applications due to their excellent mechanical properties.\cite{karger1994polypropylene,karger2012polypropylene,nomura1993ipn,jin2014effects,shen2021unexpected} Particularly in recent years, in the application of large structural materials like car bumpers, instrument panels, and door panels, the trend of non-coating these molded products has increased, making it necessary for the molded products to have a good surface appearance.\cite{nomura1993structure,yokoi2000visualization,tanaka2001phenomenon,patham2005flow,hirano2007striped,hirano2008striped,li2021experimental,liu2022multifunctional,liu2019remarkably,zhang2021facile}
	
	Flow marks are one of the representative surface appearance defects of injection-molded parts in various polymer materials, not limited to PP-type blends (including pure PP).\cite{yoshii1992influence,chang1996surface,hamada1996correlation,hobbs1996development,heuzey1997occurrence,lacrampe2000defects,patham2005flow,an2008experimental,hirano2007morphological,mathieu2022flow,owada2016visualization,gahleitner2008optical,hirano2008striped,kuroda2020effect,hirano2007striped,jeong2015flow,iannuzzi2010rheological} They are surface defects characterized by wave streaks. The streaks are roughly perpendicular to the direction of melt flow and form on the surface of the injection-molded device.\cite{yoshii1992influence} Flow marks can be classified as either same-phase type flow marks, where glossy and cloudy parts alternate and appear at the same position on the surface and inside of the molded part, or different-phase type flow marks, where the glossy and cloudy parts appear at different phases on the surface and inside. Flow marks are closely related to the unstable flow of the melt.\cite{hu2023development,grillet2002numerical,bogaerds2004stability}	
	
	Generally, it has been said that flow marks do not appear easily in materials with large die swells, but there are some materials in which flow marks do not appear easily even if the die swell is small. For this reason, the elimination of flow marks in injection-molded products has conventionally been addressed by changing injection molding conditions such as injection temperature and speed, mold temperature, etc.\cite{tanaka2001phenomenon} or molecular characteristics such as the molecular weight of the material and its distribution through repeated trial and error. In recent years, in order to break away from the flow-mark elimination measures relying on empirical methods, observations of resin flow using visualization molds have been actively conducted to elucidate the flow-mark generation mechanism.\cite{murata1993visual,yokoi2002visualization} These studies have shown that flow-mark generation is related to the unstable flow phenomenon at the material flow front during in-mold flow.\cite{yokoi2000visualization} The flow mark generation is caused by the unstable flow of the material in the mold, which is a phenomenon known as the "flow front" phenomenon. However, no clear conclusion has yet been reached on the mechanism of flow mark generation and how it leads to flow mark formation.
	
	In order to investigate the formation mechanism of flow marks, we want to visualize the melt flow process that generates the formation of flow marks and at the same time obtain all the information about the flow field, the temperature field, and the polymer conformation and their evolution. The melt flow process is microscopic for experiments, making it difficult to observe without influencing the flow, and it is also difficult to obtain microscopic information about the evolution process. For mesoscopic simulations, such as finite element analysis, information on polymer chain conformation and crystallization is lost. For molecular dynamics simulations, we can obtain this information, but the system is too macroscopic, resulting in an unmatched amount of calculations. Therefore, we would like to scale down the flow field equivalently (while ensuring it is large enough for the polymer chain size) and then visualize the flow process by means of molecular dynamics simulation and get all the information about the flow field, the temperature field, and the polymer conformation and their evolution.
	
	Through dimensional analysis, we find the dimensionless quantities that control the geometry of the flow field. These quantities can be categorized as follows: geometric dimensionless quantities related to the shape of the mold, material dimensionless quantities related to the melt and mold materials, and physical dimensionless quantities related to the flow. When the geometry of the mold changes proportionally, the melt and mold material are fixed, and the initial temperature of the melt and mold is fixed, the geometric dimensionless quantities, the material dimensionless quantities, and a part of the physical dimensionless quantities will remain constant, and the geometry of the flow field will be controlled only by the Weissenberg number $Wi$. If $Wi$ is kept constant, changing the injection speed, changing the relaxation time of the polypropylene melt, or scaling the mold will result in similar geometric shapes of the flow field. Theoretically following the above conclusions, we can scale the flow field in equal proportions, and then visualize the corresponding melt flow process, which generates flow marks, through molecular dynamics simulation. However, before doing so, we need to verify the correctness of the above conclusions. Since the dimensionless equations represent similar systems of all sizes, this paper verifies the above conclusions through a microscopic molecular dynamics simulation with the same magnitude of dimensionless quantities as the experimental injection molding process (the specific values are slightly different). The simulation results demonstrate that flow fields of the same $Wi$ systems are similar, and the flow fields of different $Wi$ systems are obviously different. The simulation results verify the conclusion of dimensional analysis to some extent.

	\section{Theory and Methods}
	
	\subsection{Dimensional Analysis Theory}
	
	Select the characteristic quantity that has a significant impact on the system as the unit, $v$, $\xi$, $\tau$, and the form before and after dimensionless transformation is as follows
	\begin{equation}\label{eqn:main}
		\lambda=f\left(v, \xi, \tau, Z, w, L, R, k T, k T^{\prime}, \varepsilon_{L J}, \varepsilon_{B}, \varepsilon_{A}, \varepsilon_{D}, m, \xi^{\prime}, t, N, n\right)
	\end{equation}
	\begin{equation}\label{eqn:dimensionless}	
		\frac{\lambda}{v \tau}=f\left(1, 1, 1, \frac{Z}{v \tau}, \frac{w}{v \tau}, \frac{L}{v \tau}, \frac{R}{v \tau}, \frac{k T}{v^{2} \xi \tau}, \frac{k T^{\prime}}{v^{2} \xi \tau}, \frac{\varepsilon_{L J}}{v^{2} \xi \tau}, \frac{\varepsilon_{B}}{v^{2} \xi \tau}, \frac{\varepsilon_{A}}{v^{2} \xi \tau}, \frac{\varepsilon_{D}}{v^{2} \xi \tau}, \frac{m}{\xi \tau}, \frac{\xi^{\prime}}{\xi}, \frac{t}{\tau}, N, n\right)
	\end{equation}	
	where the dependent variable $\lambda$ is the tiger stripe period or induced length (the dimensions are the same, the dimensional analysis is equivalent), $v$ is the melt flow rate, $\xi$ is the friction coefficient of the melt Kuhn unit, $\tau$ is the polymer chain relaxation time, $Z$ is the spacing between the upper and lower plates. $w$ is the size of the injection port, $L$ is the total length of the mold, $R$ is the Kuhn unit size, $kT$ is the Boltzmann constant and system temperature, $kT^{\prime}$ is the Boltzmann constant and mould temperature, $\varepsilon_{LJ}$, $\varepsilon_{B}$, $\varepsilon_{A}$, $\varepsilon_{D}$ are the interactions between atoms, bond energy, bond angle energy, and dihedral angle energy, $m$ is the element mass, $\xi^{\prime}$ is the friction coefficient of the melt Kuhn unit moving on the mold surface, $t$ is the observation time, $N$ is the number of chain elements, and $n$ is the number of chains. Among them, the number of independent variables needs to be equal to the degrees of freedom of the system. So, we need to eliminate some independent variables that are not independent ($t$, $N$, and $n$ have been determined by the other variables)
	\begin{equation}\label{eqn:}
		\frac{\lambda}{v \tau}=f\left(\frac{k T}{v^{2} \xi \tau}, \frac{k T^{\prime}}{v^{2} \xi \tau}, \frac{\varepsilon_{L J}}{v^{2} \xi \tau}, \frac{\varepsilon_{B}}{v^{2} \xi \tau}, \frac{\varepsilon_{A}}{v^{2} \xi \tau}, \frac{\varepsilon_{D}}{v^{2} \xi \tau}, \frac{Z}{v \tau}, \frac{w}{v \tau}, \frac{L}{v \tau}, \frac{R}{v \tau}, \frac{m}{\xi \tau}, \frac{\xi^{\prime}}{\xi}\right)
	\end{equation}
	According to the nature of dimensional analysis, since the function form is indefinite, we can combine the multiplication or division of the dimensionless independent variables to separate the pure geometric dimensionless quantities, the pure material dimensionless quantities and the physical dimensionless quantities. Sort them and separate them with a semicolon
	\begin{equation}\label{eqn:}
		\frac{\lambda}{v \tau}=f\left(\frac{w}{Z}, \frac{L}{Z}, \frac{R}{Z} ; \frac{\varepsilon_{L J}}{\varepsilon_{B}}, \frac{\varepsilon_{A}}{\varepsilon_{B}}, \frac{\varepsilon_{D}}{\varepsilon_{B}}, \frac{\xi^{\prime}}{\xi} ; W i, \frac{k T}{v^{2} \xi \tau}, \frac{\varepsilon_{B}}{v^{2} \xi \tau}, \frac{m}{\xi \tau}, \frac{T^{\prime}}{T}\right)
    \end{equation}	
	If the dimensionless independent variable in the above equation is equal to that in the macroscopic experiment, the scaling simulation can be carried out. Where
	\begin{equation}\label{eqn:}
		W i=\tau \dot{\gamma}=\frac{2 v \tau}{Z}
	\end{equation}
	Next, invariants or small quantities in production will be eliminated. In the same experiment, the mold geometry is the same, the related energy of melt polypropylene is the same, and the friction coefficient of the melt and mold surface is the same. Therefore, after eliminating the invariants, the equation becomes
	\begin{equation}\label{eqn:physical dimensionless}
		\frac{\lambda}{v \tau}=f\left(W i, \frac{k T}{v^{2} \xi \tau}, \frac{\varepsilon_{B}}{v^{2} \xi \tau}, \frac{m}{\xi \tau}, \frac{T^{\prime}}{T}\right) \\
	\end{equation}
	If we only consider the macroscopic flow field, then the above dimensional analysis is sufficient. However, if we consider the information of polymer chain orientation, then we need to consider the microscopic physical process.
	
    Observe the denominator of the second independent variable. It can be written as $v \xi \times v \tau$. The left side of the multiplication sign can be regarded as the magnitude of the friction of a Kuhn unit subjected to melt shear, while the right side of the multiplication sign can be regarded as the distance traveled in relaxation time. Therefore, the denominator can be regarded as the work done by friction on the melt in relaxation time.

    \begin{figure*}
    	\centering
    	\includegraphics[width=8cm]{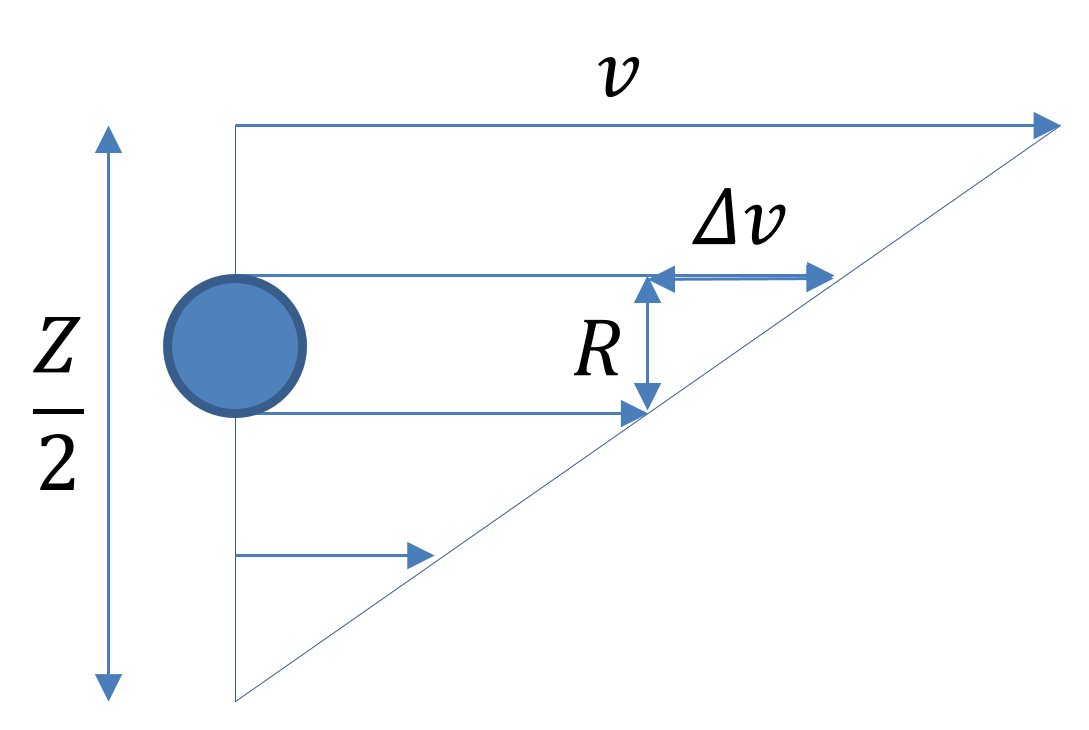}	
    	\caption{Schematic diagram of the melt unit being sheared by the flow field.}
    	\label{fig:shear}
    \end{figure*}
        
    Since the numerator of the fractional equation is the thermal kinetic energy of a Kuhn unit, the comparison of magnitudes should also correspond to the work done by friction on a Kuhn unit. As shown in Fig.~\ref{fig:shear}, the velocity of relative sliding between the units is $\Delta v=\frac{2 R}{Z} v$, and so the above equation should be transformed into
    \begin{equation}
    	\frac{k T}{\Delta v^{2} \xi \tau}=\left(\frac{Z}{R}\right)^{2} \times \frac{k T}{v^{2} \xi \tau}
    \end{equation}
    Do the same for $\frac{\varepsilon_{L J}}{v^{2} \xi \tau}$. Consider the entanglement effect~\cite{brochard1992shear}
    \begin{equation}
    \xi_{chain} = \xi \times \frac{N^3}{N_e^2}
    \end{equation}         
    After correcting each dimensionless quantity, we observe the magnitude again. In the experiments, the length of the polypropylene tangle is $N_e=32$. The number of Kuhn units and the friction coefficient in the case of Ref.~\cite{zatloukal2021reduction} are $N=1650$ and $\xi=7 \times 10^{-12} \mathrm{Ns} / \mathrm{m}$, respectively. See Table \ref{sys_para1} for the other parameters. Therefore,
    \begin{equation}\label{eqn:kt}
    \frac{N_e^2}{N^3} \times\left(\frac{Z}{2 R}\right)^2 \times \frac{k T}{v^2 \xi \tau} \approx \frac{32^2}{1650^3} \times\left(\frac{10^{-3}}{2 \times 10^{-9}}\right)^2 \times \frac{10^{-23} \times 5 \times 10^2}{\left(10^{-1} \sim 10^{-2}\right)^2 \times 7 \times 10^{-12} \times 10^{-1}} \approx 10^1 \sim 10^{-1}
    \end{equation}   
    According to Ref.~\cite{wei2018effect}, $\frac{\varepsilon_{L J}}{k T}=10^{-1}$
    \begin{equation}
    \frac{N_e^2}{N^3} \times\left(\frac{Z}{2 R}\right)^2 \times \frac{\varepsilon_{L J}}{v^2 \xi \tau} \approx 1 \sim 10 ^ {-2}
    \end{equation} 
    Kuhn unit mass $m=0.1878 \mathrm{~kg} / \mathrm{mol}$.\cite{zatloukal2021reduction}
    \begin{equation}
    \frac{N_e^2}{N^3} \times \frac{m}{\xi \tau} \approx \frac{32^2}{1650^3} \times \frac{0.1878 \div\left(6.02 \times 10^{23}\right)}{7 \times 10^{-12} \times 10^{-1}} \approx 10^{-19}
    \end{equation}  
    Melt temperature $T=483.15 \mathrm{~K}$ and mold temperature $T=323.15 \mathrm{~K}$. \cite{maeda2007flow}
    \begin{equation}
    \frac{T^{\prime}}{T}=\frac{323.15}{483.15} \approx 0.7
    \end{equation} 

    The dimensionless number of mass is very small and can be eliminated. Using the Doi-Edwards crawling model,\cite{doi1996introduction,doi1988theory} we can obtain the relaxation time
    \begin{equation}\label{eqn:Doi-Edwards}
	\tau \approx \frac{\xi b^{2}}{k T} \times \frac{N^{3}}{N_e}
    \end{equation}	 
    by replacing Eq.~(\ref{eqn:kt}) with the relaxation time
    \begin{equation}
	\frac{N_e^2}{N^3} \times\left(\frac{Z}{2 R}\right)^2 \times \frac{k T}{v^2 \xi \tau}=N_e \times\left(\frac{Z}{2 v \tau}\right)^2=\frac{N_e}{W i^2}
    \end{equation}    
    The numerator and denominator have been determined by other quantities, and this independent variable is equivalent to having been determined, and therefore can be eliminated. Divide the independent variables, eliminate the two dimensionless numbers, and then multiply both sides of Eq.~(\ref{eqn:physical dimensionless}) by $Wi$
    \begin{equation}
    \frac{\lambda}{Z}=f\left(W i, \frac{\varepsilon_{L J}}{k T}, \frac{T^{\prime}}{T}\right)
    \end{equation}     
    Since $\lambda$ represents an arbitrary length measure of the flow field, if $\frac{\lambda}{z}$ remains constant for different systems, then it indicates that the different systems are geometrically similar. Therefore, we can conclude that when the geometry of the mold changes in equal proportions, while the melt and mold material and the initial temperature of the melt and mold remain fixed, the geometry of the flow field will be solely controlled by the Weissenberg number $Wi$. If $Wi$ is kept constant, changing the injection speed, changing the relaxation time of the polypropylene melt, or scaling the mold will result in similar geometric shapes of the flow field. If the size of the mold is not changed, the geometric shape of the flow field will be the same. Since the dimensionless equations represent similar systems of all sizes, we will then verify the above conclusions through molecular dynamics simulations at a smaller scale.

	\subsection{Molecular Dynamics Simulation}
	
	The polymer consists of a bead-spring chain with a diameter of $\sigma$ and a degree of polymerization of $\mathrm{N}$ (see Table \ref{sys_para1}). The metal atoms are composed of beads with a diameter of $\sigma$, and the interactions between all the beads in the system are modeled using the truncated-shifted Lennard-Jones (LJ) potential
		\begin{small}
		\begin{equation}\label{eqn:lj}
			U_\mr{LJ}(\mb r)\!=\!\left\{\!\begin{array}{ll}
				4\varepsilon_\mr{\scriptscriptstyle LJ}\!\left[\left(\!\frac{\sigma}{r}\!\right)^{12}-\left(\!\frac{\sigma}{r}\!\right)^{6}-\left(\!\frac{\sigma}{r_\mr{cut}}\!\right)^{12}+\left(\!\frac{\sigma}{r_\mr{cut}}\!\right)^{6}\right], & \!  r \leq r_\mr{cut} \\
				\! 0, &\!  r>r_\mr{cut}
			\end{array}\right.
		\end{equation}
	\end{small}	
    The cutoff distance for the bead-by-bead interaction $r_{c u t}=2.5 \sigma$ . Referring to Ref. 5, we take the Lennard-Jones interaction parameter for the intermelt interaction to be $\varepsilon_{L J}=0.1 k_B T$, where $k_B$ is Boltzmann's constant and $T$ is the absolute temperature. For metal-polymer interactions, if we want the flow field to be similar to the experimental one, then $\frac{\xi^{\prime}}{\xi}$ needs to be equal to the experimental one. However, in order to verify the correctness of the dimension analysis, the exact interaction size is not needed, only the simulation and the system that the dimensionless quantity is equal in magnitude, so in order to prevent penetration, the melt-metal interaction is taken as $\varepsilon_{L J}=2.0 k_B T$, which has been tested $\frac{\xi^{\prime}}{\xi} \approx 20$. In addition, to simplify the model we control that the metal atoms are stationary and there is no interaction between the metal atoms.

	Beads are bonded to form polymer chains, and all bonds in the system are represented by harmonic potentials:
	\begin{equation}\label{eqn:harm}
	    U_{\text {harm }}=K\left(r-r_{0}\right)^{2}
    \end{equation}	
	where $k=300 k_B T / \sigma^2$ is the spring constant and $r_0=1.2 \sigma$ is the bond length.
	
	For simplicity, we first validate the conclusions of the dimensional analysis at the same melt temperature as the mold temperature. Therefore, we simulate it in the NVT ensemble, where a constant temperature is maintained by coupling with a Langevin thermostat. In this case, the motion equation of the $i$th bead is
	\begin{equation}\label{eqn:langevin}
		m \frac{\mathrm{d} \vec{v}_i(t)}{\mathrm{d} t}=\vec{F}_i(t)-\xi_l \vec{v}_i(t)+\vec{F}_i^R(t)
	\end{equation}	
	Where $m$ represents the mass of the bead, and in our system, the mass of all beads is set to be uniform, $\vec{v}_i(t)$ denotes the bead velocity, and $\vec{F}_i(t)$ represents the net deterministic force acting on the $i$th bead. The stochastic force $\vec{F}_i^R(t)$ has a zero mean value $\big\langle\vec{F}_i^R(t)\big\rangle=0$, and has a $\delta$-functional correlations, $\big\langle\vec{F}_i^R(t) \vec{F}_j^R\left(t^{\prime}\right)\big\rangle=6 \xi_l k_B T \delta_{i j} \delta\left(t-t^{\prime}\right)$. where $\tau_\mr{\scriptscriptstyle LJ}=\sigma(m/\varepsilon_\mr{\scriptscriptstyle LJ})^{1/2}$ is the standard LJ time. The velocity-Verlet algorithm is utilized to numerically integrate the motion equation with a time step $\Delta t=0.005\tau_\mr{\scriptscriptstyle LJ}$. All molecular dynamics simulations are performed via the LAMMPS software package \cite{1995Fast}. The unit friction coefficient $\xi$ is the superposition of the friction coefficient $\xi_l$ in the Langevin thermostat and the friction coefficient $\xi_{c}$ of units colliding with each other. Therefore, by regulating the friction coefficient of the Langevin thermostat we can regulate the friction coefficient of the melt unit. We chose four different friction coefficients $\xi_l$ is $40 m / \tau_{\mathrm{LJ}}$, $26.7 m / \tau_{\mathrm{LJ}}$, $17.8 m / \tau_{\mathrm{LJ}}$ and $7.9 m / \tau_{\mathrm{LJ}}$, corresponding to $\xi$ of $23.2 m / \tau_{\mathrm{LJ}}$, $16.7 m / \tau_{\mathrm{LJ}}$, $11.6 m / \tau_{\mathrm{LJ}}$, and $6.2 m / \tau_{\mathrm{LJ}}$ when the particle number density is 0.85. The value of $\xi_l$ shrinks by $1.5$ times, $2.25$ times, and $5.06$ times, respectively, and $\xi$ is approximately proportional to $\xi_l$.

\begin{figure*}
		\centering
		\includegraphics[width=15cm]{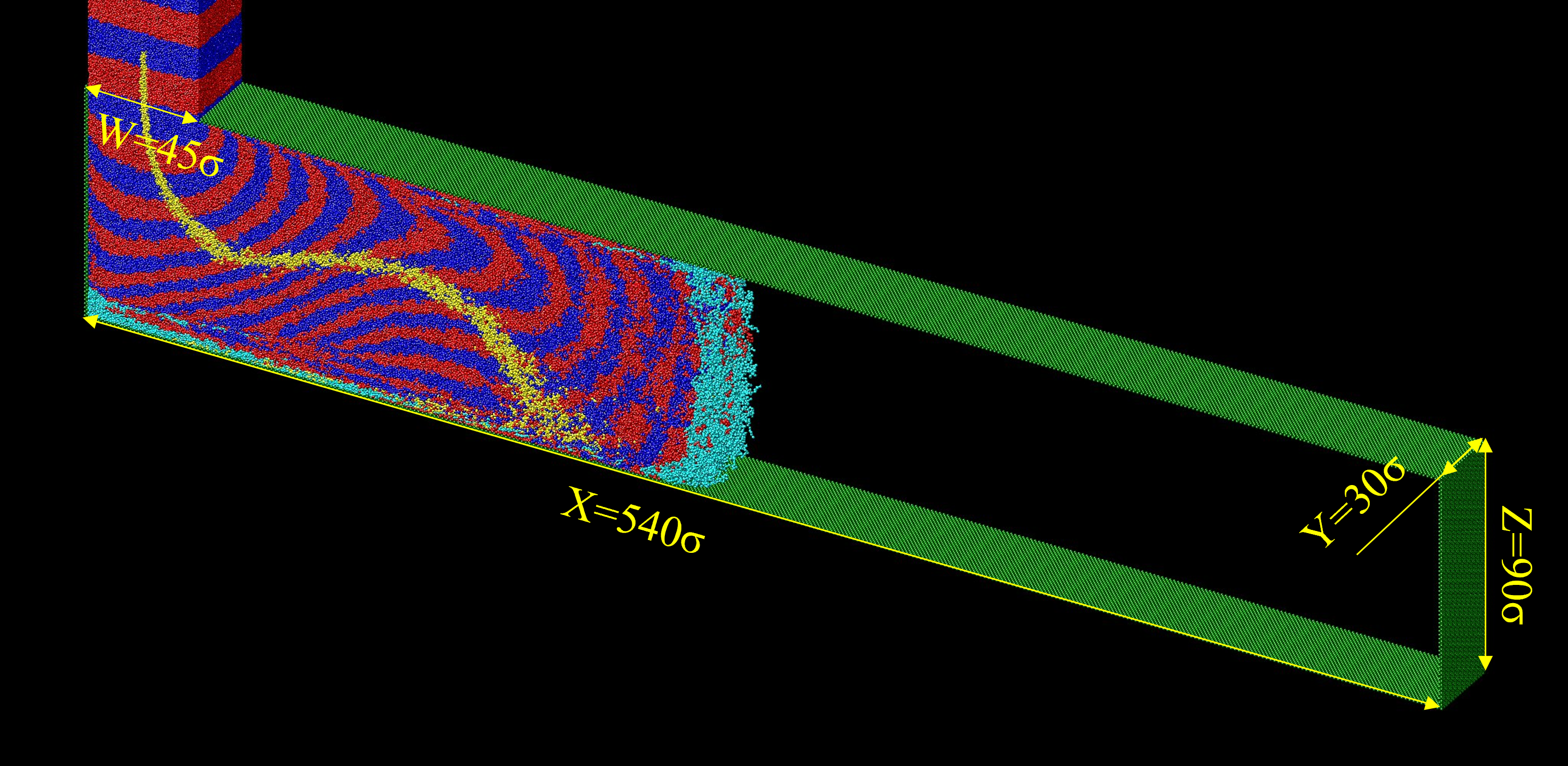}	
		\caption{Structure and size of injection molded model in our simulation.}
		\label{fig:injection}
	\end{figure*}

    Our injection mold is shown in Fig.~\ref{fig:injection}, where the polymer melt is injected through a thin tube on the left side, and a plate at the end is pushed downward at a constant speed so that the melt flows between the two metal plates. The green color represents metal atoms, and the red and blue colors are the same polymer melt. The red and blue melt flow successively from the thin tube into the two plates, so the dividing line between them can be considered isochronous. Additionally, the atoms flowing through the center of the thin tube are marked yellow, and a periodic curve is formed after flowing. The light green color at the bottom represents the front of flow during initial injection. The $x$-direction and $z$-direction are fixed boundaries, and the $y$-direction is a periodic boundary. According to the calculated quantities, we chose three system sizes (different systems satisfying geometrical similarity) with two-plate spacings $Z$ of $40 \sigma$, $60 \sigma$, and $90 \sigma$ (the simulated system is a free-connecting chain, and we would like to have one unit corresponding to one polypropylene Kuhn unit, which has a size of $11$ \AA, so the two-plate spacing corresponds to the real system's approximate size of $400$ \AA, $600$ \AA, and $900$ \AA). The maximum system geometric parameters are shown in Fig.~\ref{fig:injection}, the length of the vertical pipe part $Z^{\prime}=450 \sigma$, the number of melt particles is $1,541,788$, and the number of metal particles is $58,444$. The parameters of the other size systems are scaled proportionally.
    
    For the simulated system, the relaxation time $\tau$ of the polymer itself is not clearly defined. However, the system needs to satisfy the condition that each dimensionless independent variable is equal to that of the macroscopic experiment or is a very large or very small quantity, as in the macroscopic experiment. Hence, it is necessary to determine the relaxation time $\tau$ of the polymer in advance. Using Eq.~(\ref{eqn:Doi-Edwards}), we can calculate the relaxation time $\tau$ for different polymerization degrees $N$. For systems where we need to control the relaxation time, we can work backwards to calculate the degree of polymerization we need. To improve the accuracy of selecting $N$, we also measure the relaxation times corresponding to different $N$ by the unit mean square displacements versus time predicted by the crawling model. We use the $N$ computed in the previous step as a baseline and increase or decrease $N$ until we find the $N$ that is closest to the target relaxation time. The relationship between the unit mean square displacement and time predicted by the crawling model is divided into four regions based on different power relationships, with respective exponents of 1/2, 1/4, 1/2, and 1.
    The turning point in the third and fourth regions is the crawling time $\tau_{\text {rep}}$, which is the relaxation time we need. We can find this turning point by fitting, and thus determine the corresponding relaxation time. For the measurement of mean square displacements, we choose a cube with a side length of $60 \sigma$ as the simulation box, while the particle number density is $0.85 \sigma^{-3}$. For different degrees of polymerization, the particle number density is kept constant and $\xi_l$ is set to $1 m / \tau_{\mathrm{LJ}}$, corresponding to $\xi$ being equal to $2.1 m / \tau_{\mathrm{LJ}}$, and all other parameter selections are the same as before. According to Ref.\cite{schulz1993reptation}, the central unit of the chain was selected for measurement to improve accuracy.

\begin{figure*}
    	\centering
    	\includegraphics[width=15cm]{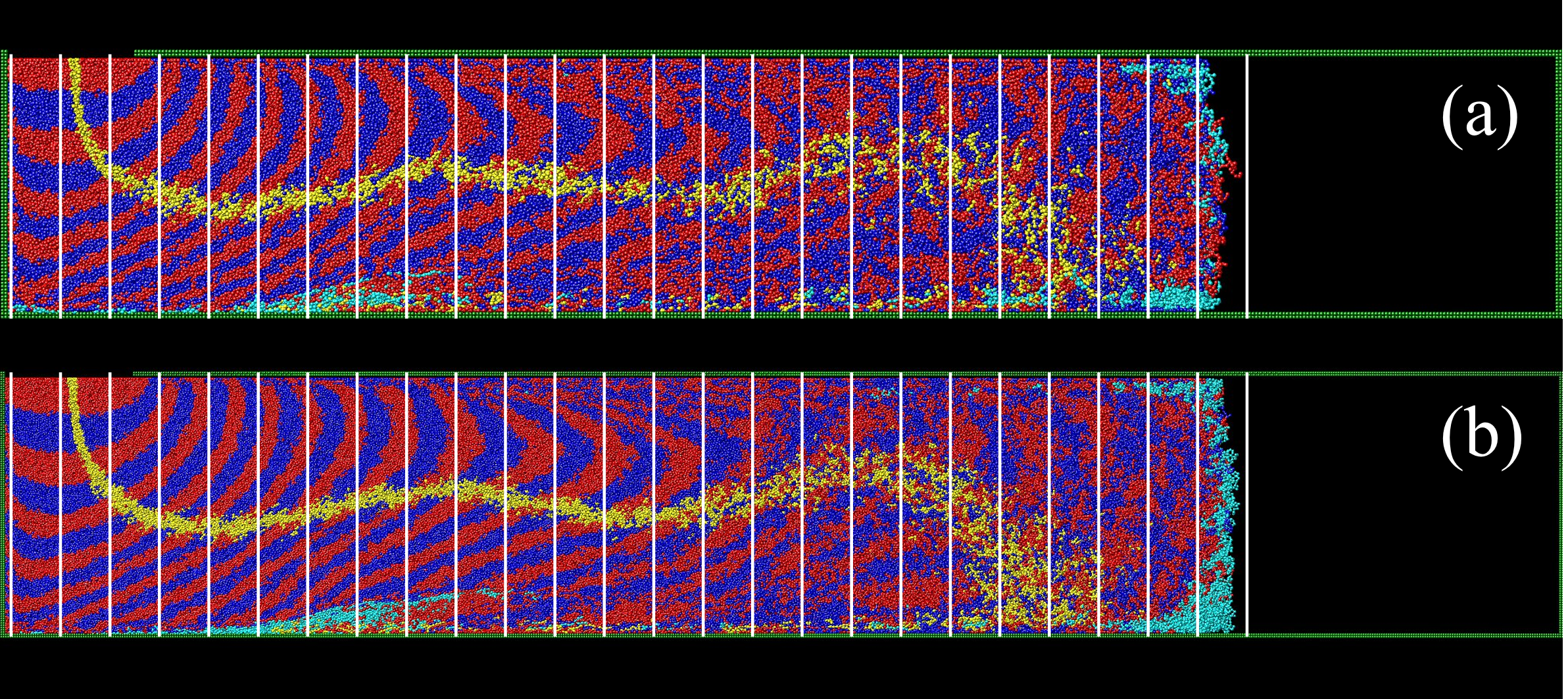}	
    	\caption{Demonstration of system division. Above is the system with $Z=60 \sigma$ (a), below is the system with $Z=90 \sigma$ (b). This is a schematic diagram, the actual division is more dense.}
    	\label{fig:similar}
    \end{figure*}

    In order to assess the degree of similarity or dissimilarity of the system, we quantified the system similarity $S$. As shown in Fig.~\ref{fig:similar}, we divided the system into 240 chunks, of which the number of chunks with yellow atoms is $n$ (Fig.~\ref{fig:similar} is schematic, the actual division is more dense). The coordinates of the $z$ direction of the center of mass of the yellow atoms in each block are calculated. The larger of the two compared systems is scaled down by a factor of $k$ so that the two systems are the same size (some adjustments are needed for the different starting and ending positions of the yellow atoms in the $x$-direction of the different systems). Then, calculate the root mean square error $\Delta$ of the coordinates in the $z$-direction of the two systems, and we define the similarity $S=\frac{1}{1+a * \Delta}$. To keep the similarity within a suitable interval, we choose $a=0.8$.
    \begin{equation}\label{eqn:similarity}
    	\Delta=\sqrt{\frac{\sum_i^n\left(z_i^{\prime} / k-z_i\right)^2}{n}}
    \end{equation}

	\section{Results and discussion}
	
\begin{figure*}
		\centering
		\includegraphics[width=15cm]{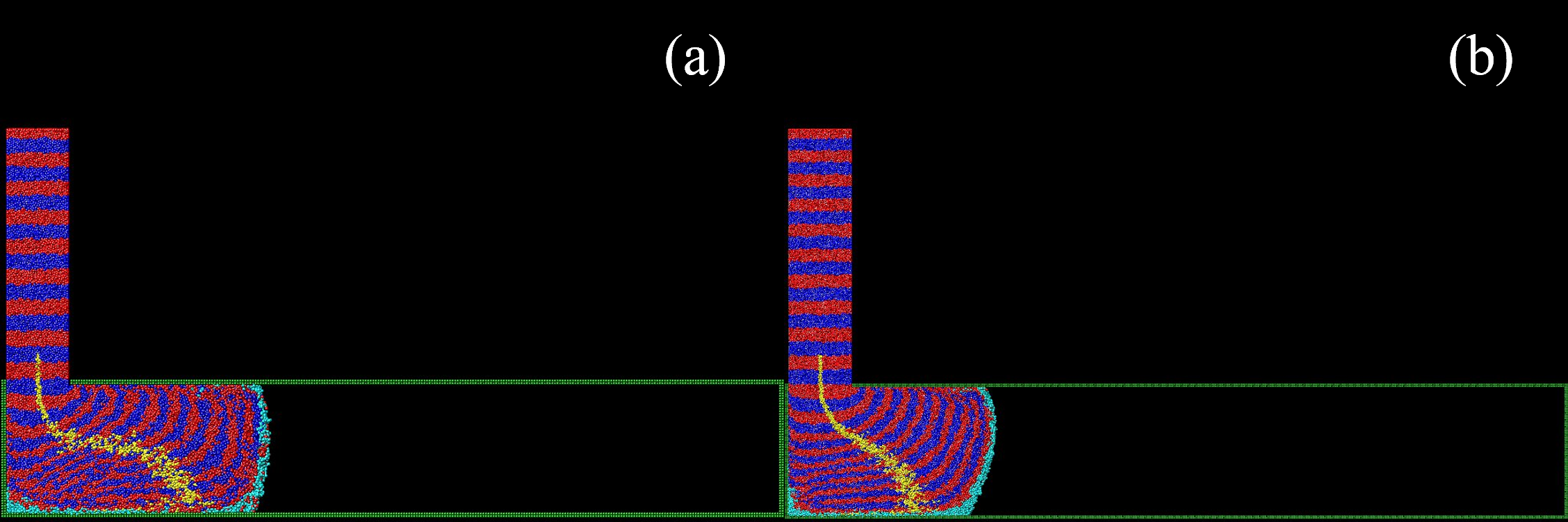}	
		\caption{When we satisfy only the equality of $Wi$, but not the equality of $kT/v^{2}\xi t$, the velocity of the plate push is scaled equivalently, but the flow rate at the pipe outlet is not scaled equivalently. The left side shows the system with $Z=40 \sigma$, and the right side shows the system with $Z=60 \sigma$.}
		\label{fig:tube}
	\end{figure*}

\begin{figure*}[h]
		\centering
		\includegraphics[width=15cm]{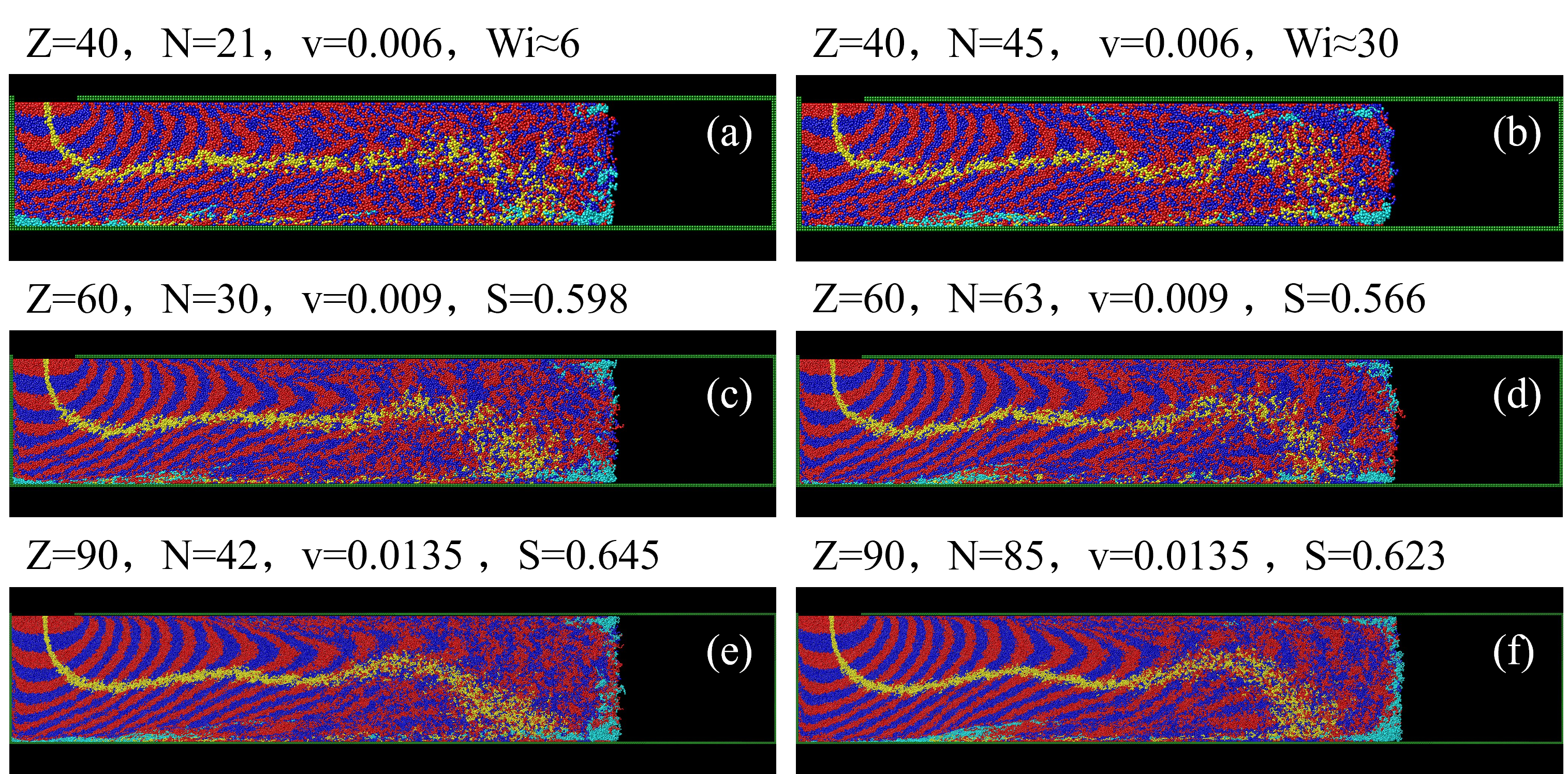}	
		\caption{The systems where $v$ and $Z$ are proportional: (a)(c)(e) for the system $Wi \approx 6$, (b)(d)(f) for the system $Wi \approx 30$. From top to bottom, $\xi_l$ is $40 m / \tau_{\mathrm{LJ}}$ for (a)(b), $17.8 m / \tau_{\mathrm{LJ}}$ for (c)(d), and $7.9 m / \tau_{\mathrm{LJ}}$ for (e)(f), respectively.}
		\label{fig:vz}
	\end{figure*}

	Before verifying whether the flow field is controlled by $Wi$ through simulation, we must first do some preparatory work. First, we consider that since there is a vertical pipe section, it is necessary to match not only the dimensionless quantity $Wi$ between the two plates but also the dimensionless quantity $k T/v^{2} \xi t$ inside the pipe (which can be obtained by a similar dimensional analysis). Otherwise, for systems of different sizes, even if the invisible plate push velocity is scaled equally, the pipe outlet flow rate does not satisfy the scaling, which leads to the flow rate of the melt between the two plates not satisfying the scaling, as shown in Figs.\ref{fig:tube}(a) and (b). In the macro system, because the size Z is large, so $v$ and $t$ are large, the above dimensionless quantity is very small and can be ignored, so the scaling of the macro system should not consider this dimensionless quantity, but the volume reduced simulation system needs to consider it. From Einstein's formula $D=k T / \xi$:
	\begin{equation}\label{eqn:Einstein}
		\frac{k T}{v^{2} \xi t}=\frac{D}{v^{2} t}=\frac{D t}{v^{2} t^{2}} \approx \frac{\left\langle[r(t)-r(0)]^{2}\right\rangle}{(v t)^{2}}
	\end{equation}	
    It can be seen that this dimensionless quantity represents the ratio of the average distance moved by the particle per unit time $t$ to the distance pushed by the invisible plate. For different systems, when this dimensionless quantity is equal, the average velocity of the particle movement is proportional to the velocity pushed by the invisible plate, and the flow rate of the melt between the two plates will satisfy the scaling when the velocity pushed by the invisible plate is scaled equivalently, as shown in Figs.\ref{fig:vz}(a), (c) and (e).
	
	Next we consider the proportionality of the parameters between the different systems. Since we have controlled for equal proportional changes in mold geometry, constant melt and mold materials, and constant melt and mold temperatures, the dimensionless equations are simplified to:
	\begin{equation}\label{eqn:Wi}
		\frac{\lambda}{Z}=f\left(\frac{v \tau}{Z}\right)
	\end{equation}	
	 Our goal is to prove that the flow field remains unchanged for any combination of values $v$, $\tau$, and $Z$, as long as $Wi$ remains constant. However, logically we only need to show that this conclusion holds in two cases: 1. The shape of the flow field is constant when $v$ and $Z$ are varied proportionally. 2. When $v$ and $\tau$ are varied inversely, the shape of the flow field remains constant. Because any combination of $v$, $\tau$, and $Z$ can be obtained by proportionally changing $v$ and $Z$, and inversely changing $v$ and $\tau$, as long as it satisfies the requirement that $Wi$ is invariant. Thus, if the conclusion holds in the above two cases, then it is proof that the conclusion holds in any case.			
    
    Since we need to satisfy both $\frac{L}{Z}, \frac{v \tau}{Z}$ and $\frac{k T}{v^2 \xi t}$ are unchanged, and $L=vt$. Recombining these dimensionless numbers yields $\frac{t}{\tau}, \frac{v \tau}{Z}$ and $\frac{k T}{v^2 \xi \tau}$ are unchanged, while $\tau \approx \frac{\xi b^2}{k T} \times \frac{N^3}{N e}$, so for the choice of the scaling relation of the parameters our only strategy is: 1. When $v$ and $Z$ increase proportionally, reduce $\xi$ by the square multiple of the increase in $v$ (increasing $T$ is equivalent to decreasing $\xi$), changing $N$ ensures that $\tau$ remains constant, and the program running time $t$ remains constant. 2. When $v$ and $\tau$ are varied inversely, $Z$ remains unchanged, and if $v$ decreases then $\xi$ increases in the same proportion, $\tau$ therefore increases in the same proportion, and the program running time $t$ increases in the same proportion.
    
    Take the system we simulated as an example: The two-plate spacing $Z$ for the three systems are $40 \sigma$, $60 \sigma$ and $90 \sigma$, respectively. When $v$ and $Z$ increase proportionally, $Z$ increases by 1.5, therefore $v$ increases by 1.5 and $\xi$ decreases by 2.25. In order to ensure that $\tau$ remains constant, we will increase the degree of polymerization, $N$. The exact value of the degree of polymerization will be determined in the next steps. When $v$ and $\tau$ are varied inversely, $v$ decreases by a factor of 1.5, so $\xi$ increases by a factor of 1.5, and $\tau$ naturally increases by 1.5 without changing $N$.

\begin{figure*}[t]
		\centering
		\includegraphics[width=15cm]{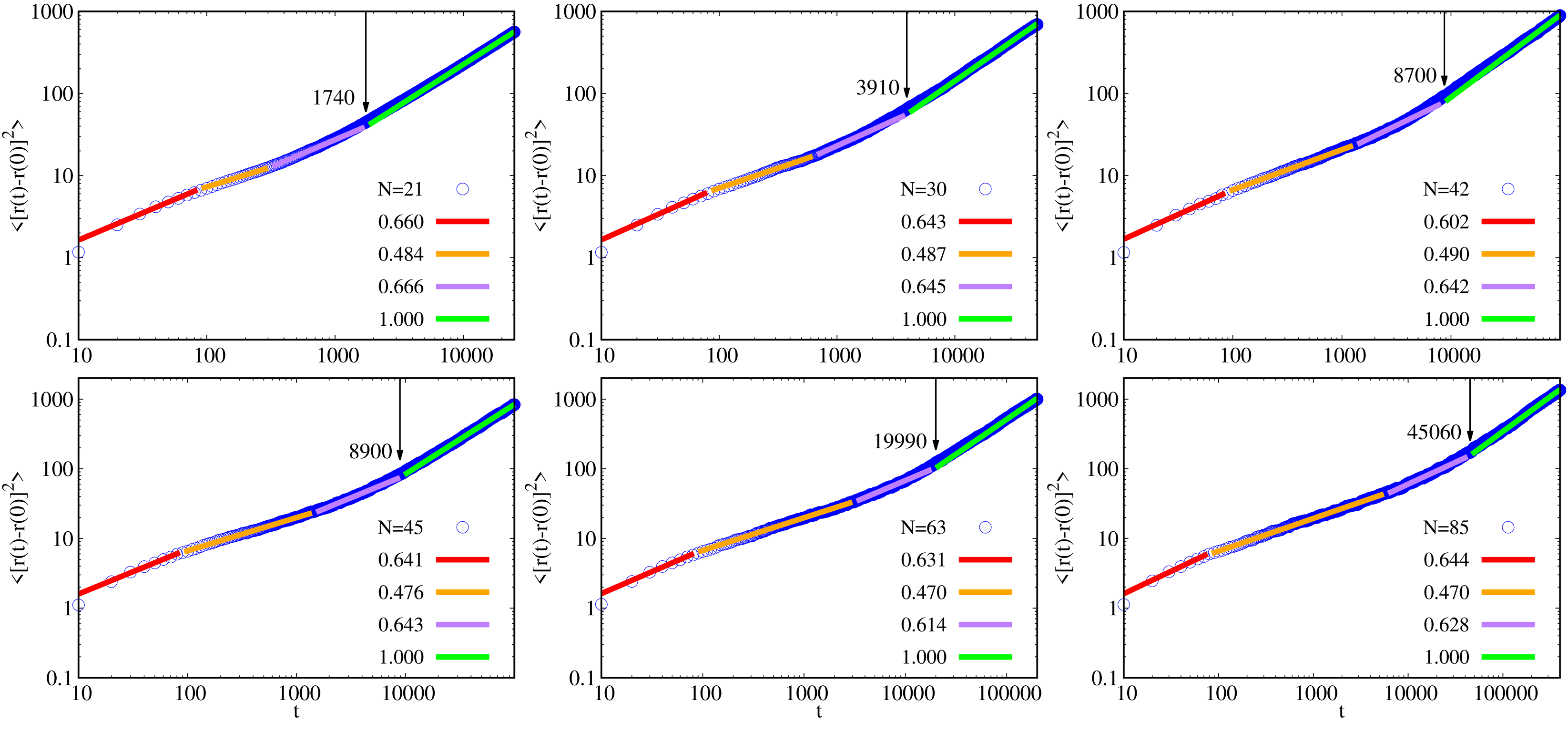}	
		\caption{The relationship between the mean square displacements of the center unit of the chain and time, in logarithmic coordinates. The system temperature is $1 k_B T$, with the langevin thermostat friction coefficient selection $\xi_l=1 m / \tau_{\mathrm{LJ}}$ and measured $\xi=2.1 m / \tau_{\mathrm{LJ}}$. The degree of polymerization for each case is as follows: (a) $N=21$, (b) $N=30$, (c) $N=42$, (d) $N=45$, (e) $N=63$, (f) $N=85$. The straight lines represent the fit to the data for each region, and the key values show their slopes.}
		\label{fig:relaxion}
	\end{figure*}

    In order to determine the size of $N$ when $v$ and $Z$ are increased in equal proportion, we first consider how much an increase in $N$ increases the relaxation time by a factor of 2.25 if the coefficient of friction remains constant, and then we choose this $N$ and reduce the coefficient of friction by a factor of 2.25 to ensure that the relaxation time $\tau$ remains unchanged in a close approximation. First, we can back-calculate using the Doi-Edwards crawling model, see Eq.~(\ref{eqn:Doi-Edwards}). For the smallest system $Z=40 \sigma$, we choose $N=21$ and $N=45$, respectively, and calculate the degrees of polymerization to be $N=28$ and $N=59$ for $Z=60 \sigma$. For $Z=90 \sigma$, the degree of polymerization is $N=37$ and $N=77$. By measuring the mean square displacement of the central unit of the polymer chain over time \cite{schulz1993reptation}, we find that these degrees of polymerization correspond to relaxation times that increase by a factor of less than 2.25. Therefore, we increase the degree of polymerization and measure the relaxation time until it increases by a factor of 2.25. As shown in Fig.~\ref{fig:relaxion}, we fit the mean-square displacement over time with four straight lines and divided four regions with different slopes based on the intersection of the lines. The slopes we obtained for each region are slightly higher than those of the crawling model, similar to the results in the Ref.\cite{schulz1993reptation}. The turning points in the third and fourth regions are the relaxation times. The relaxation time $\tau=1740 \tau_{\mathrm{LJ}}$ when the degree of polymerization $N=30$. When the relaxation time is expanded by a factor of $2.25$ to $3915 \tau_{\mathrm{LJ}}$, the closest value of the relaxation time is at $N=30$. When it is further expanded by a factor of $2.25$ to $8809 \tau_{\mathrm{LJ}}$, the closest value of the relaxation time is at $N=42$. The relaxation time is $\tau = 8900 \tau_ {\mathrm {LJ}}$ when the degree of polymerization is $N = 45$. When the relaxation time is expanded by a factor of $2.25$ to $20025 \tau_{\mathrm {LJ}}$, the closest value of the relaxation time is at $N=63$. When it is further expanded by a factor of $2.25$ to $45056 \tau_{\mathrm{LJ}}$, the closest value of the relaxation time is at $N=85$. Ultimately, we determined that for $Z=60 \sigma$, the degrees of polymerization are  $N=42$ and $N=63$. For $Z = 90 \sigma$, the degrees of polymerization are $N = 59$ and $N = 85$.

	After considering the entanglement effect and adding the dimensionless quantity in the vertical pipe, the complete dimensionless equation is:
	\begin{equation}\label{eqn:complete}
    \frac{\lambda}{Z}=f\left(\frac{w}{Z}, \frac{L}{Z}, \frac{R}{Z} ; \frac{\varepsilon_A}{\varepsilon_B}, \frac{\varepsilon_D}{\varepsilon_B}, \frac{\xi^{\prime}}{\xi} ; W i, \frac{\varepsilon_{L J}}{k T}, \frac{\varepsilon_B}{k T}, \frac{k T}{v^2 \xi \tau}, \frac{N_e^2 m}{N^3 \xi \tau}\right)
	\end{equation}
    With the above preparations, we are now able to determine the full parameters of the systems. By choosing the parameters, we want to make each dimensionless independent variable in Eq.~(\ref{eqn:complete}) in the simulated system equal to that of the macroscopic experiment or be a very large or very small quantity as in the macroscopic experiment. The specific parameters are shown in Table \ref{sys_para1}. In Table \ref{sys_para1}, we compare the two simulated systems with $Z=40 \sigma$ to the experiments, and the experimental parameters are taken from Ref.\cite{maeda2007flow,zatloukal2021reduction}. According to the above, the polymerization degree, friction coefficient, and relaxation time of the two simulation systems have been determined. To ensure that $Wi$ is close to the experimental value, we control the total number of steps for the invisible plate pushing from the top of the pipe to the bottom of the pipe at a uniform speed to be 6,000,000 (the total number of steps changes proportionally when $v$ and $\tau$ vary inversely), and the average flow velocity between the two plates is calculated to be $v \approx 0.006$. For the systems with $N = 30$ and $N = 45$, we have $\tau = \frac{23.2}{2.1} \times 1740 = 19222$ and $\tau = \frac{23.2}{2.1} \times 8900 = 98324$, respectively. From these values, we obtain $Wi = \frac{2v\tau}{Z} = 5.8$ and $Wi = \frac{2v\tau}{Z} = 29.5$. Here, $\frac{w}{Z}$ represents the ratio of the injection hole size (i.e., the size of the vertical tube) to the spacing between the plates. Horizontal injection molding has $\frac{w}{Z} = 0.5$ according to Ref.\cite{maeda2007flow}, while vertical injection molding in industrial production has $\frac{w}{Z} = 4$. Our model is based on vertical injection molding. However, in order to enhance the periodic flow characteristics to facilitate the comparison of different systems, we selecte the case of $\frac{w}{Z} = 0.5$ for simulation. For $\frac{L}{Z}$, we chose $L$ as the length at which flow marks begin to appear in the experiment, and as the total length of the injection molding in the simulation. It can be seen that the simulation requires an increase in the amount of melt in order to observe the flow corresponding to the flow marks, which will be done in our future work. As for $\frac{R}{Z}$, the experimental quantity is very small, but the simulation quantity is not sufficiently small, which will affect the similarity of the system to some extent. Referring to Ref.\cite{wei2018effect}, we can obtain $\frac{\varepsilon_{\mathrm{LJ}}}{k T}$ and $\frac{\varepsilon_B}{k T}$ since the system consists of freely connected chains, we do not consider the other interaction potentials. As for $\frac{\xi^{\prime}}{\xi}$, no experiments have been found to accurately measure this value. Therefore, we refer to Ref.\cite{brochard1992shear}. In ideal conditions, with a smooth solid surface and no chains attached to it, one expects that the friction $k$ is comparable to what it is in a fluid of monomers. In the simulation, this value is determined by the interaction between the melt and the metal. As for $\frac{k T}{v^2 \xi t}$ and $\frac{N_e^2 m}{N^3 \xi \tau}$, they can be calculated using the aforementioned parameters. $\frac{k T}{v^2 \xi t}$ is a very small quantity for experiments and not sufficiently small for simulations, while $\frac{N_e^2 m}{N^3 \xi \tau}$ is negligible in both experiments and simulations. Table \ref{sys_para2} compares the parameters of the different systems in the simulation for the two smallest simulated systems, the system after the proportional change in $v$ and $Z$, and the system after the inverse change in $v$ and $\tau$. The units of each system are LJ units.
      	
	\begin{table*}[t]
		\caption{System parameters}
		\label{sys_para1}	
        \resizebox{15cm}{!}{
		
        \begin{tabular}{|c|c|c|c|c|c|c|c|c|c|c|c|c|c|}
        	\hline system & $N$ & $\xi$ & $\tau$ & $v$ & $W i$ & $\frac{w}{Z}$ & $\frac{L}{Z}$ & $\frac{R}{Z}$ & $\frac{\varepsilon_{\mathrm{LJ}}}{k T}$ & $\frac{\varepsilon_B}{k T}$ & $\frac{\xi^{\prime}}{\xi}$ & $\frac{k T}{v^2 \xi t}$ & $\frac{N_e^2 m}{N^3 \xi \tau}$ \\
        	\hline experiment
        	 & 2000 & $10^{-12}$ & 0.1 & $0.01 \sim 0.1$ & $1 \sim 20$ & 0.5 & $40 \sim 150$ & $10^{-6}$ & 0.1 & 300 & 1 & \begin{tabular}{l}
        		$10^{-4}$ \\
        		$\sim 10^{-6}$
        	\end{tabular} & $10^{-19}$ \\
        	\hline simulation 1 & 21 & 23.2 & 19222 & 0.006 & 5.8 & 0.5 & 4.7 & $\frac{1}{40}$ & 0.1 & 300 & 10 & 0.04 & $10^{-8}$ \\
        	\hline simulation 2 & 45 & 23.2 & 98324 & 0.006 & 29.5 & 0.5 & 4.7 & $\frac{1}{40}$ & 0.1 & 300 & 10 & 0.04 & $10^{-10}$ \\
        	\hline
        \end{tabular}
        }
    \end{table*}		

	\begin{table*}[t]
		\caption{System parameters}
		\label{sys_para2}	
		\setlength{\tabcolsep}{5.5mm}
		
       	\begin{tabular}{ccccc}
       		\hline \multicolumn{3}{c}{Common parameters} & \multicolumn{1}{c}{$W i \approx 6$} & \multicolumn{1}{c}{$W i \approx 30$} \\
       		\hline $Z$ & $v$ & $\xi_l$ & $N$ & $N$ \\
       		\hline 40 & 0.006 & 40 & 30 & 45 \\
       		60 & 0.009 & 17.8 & 42 & 63 \\
       		60 & 0.006 & 26.7 & 42 & 63 \\
       		90 & 0.0135 & 7.9 & 59 & 85 \\
       		90 & 0.006 & 17.8 & 59 & 85 \\
       		\hline
       	\end{tabular}
        
	\end{table*}

	\begin{figure*}[h]
		\centering
		\includegraphics[width=15cm]{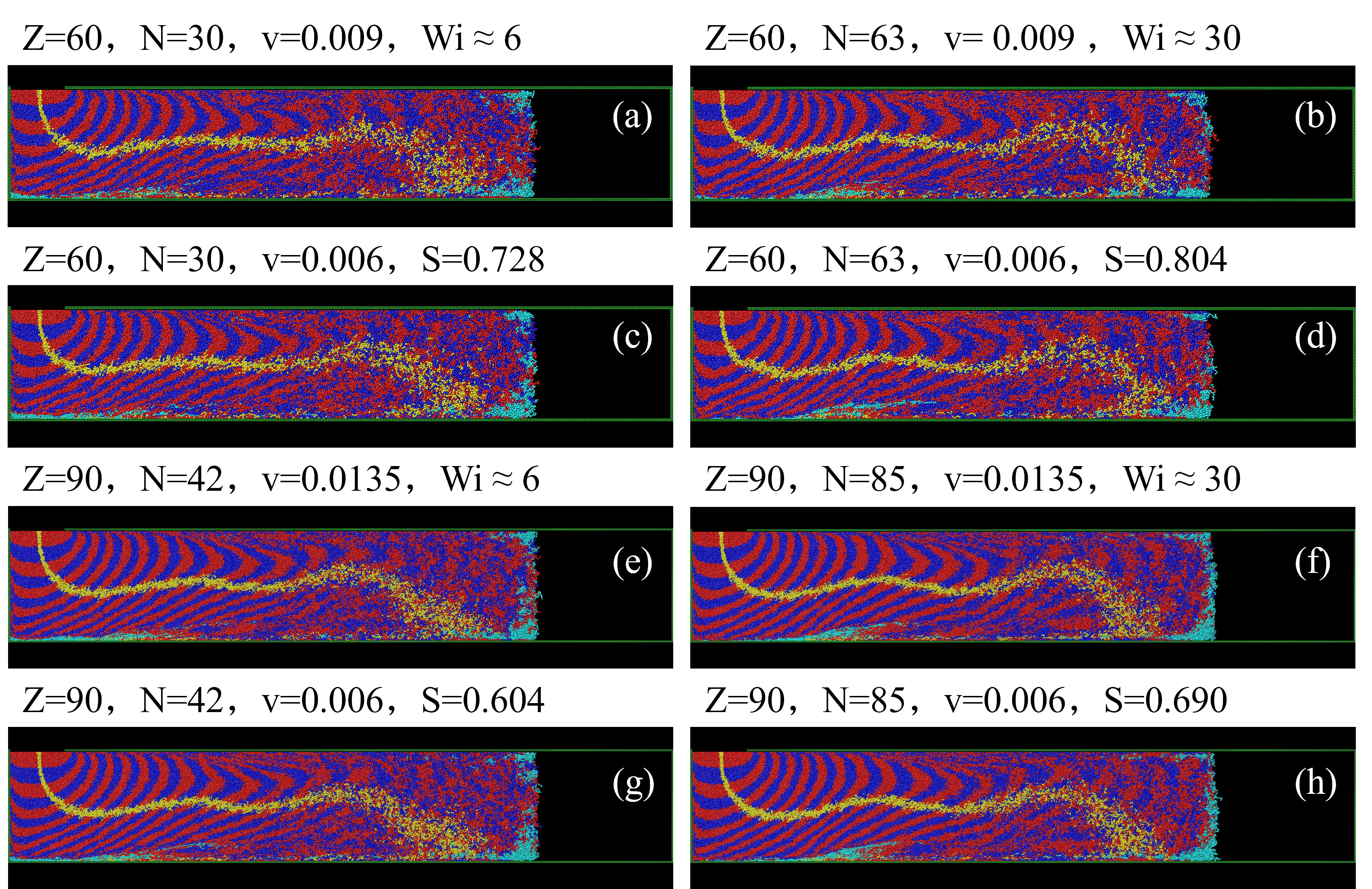}	
		\caption{The systems where $v$ and $\tau$ are inversely proportional: (a)(c)(e)(g) are the systems of $W i \approx 6$, (b)(d)(f)(h) are the systems of $W i \approx 30$. From top to bottom, $\xi_l$ is $17.8 m / \tau_{\mathrm{LJ}}$ for (a)(b), $26.7 m / \tau_{\mathrm{LJ}}$ for (c)(d), $7.9 m / \tau_{\mathrm{LJ}}$ for (e)(f), and $17.8 m / \tau_{\mathrm{LJ}}$ for (g)(h), respectively.}
		\label{fig:vt}
	\end{figure*}

	Based on the above parameter selection, we carried out molecular dynamics simulations, and the flow fields obtained are shown in Fig.~\ref{fig:vz} and Fig.~\ref{fig:vt}. Qualitatively, we can observe that the flow fields are similar for systems with the same $Wi$ and significantly different for systems with different $Wi$.
	
	\begin{figure*}[h]
		\centering
		\includegraphics[width=8cm]{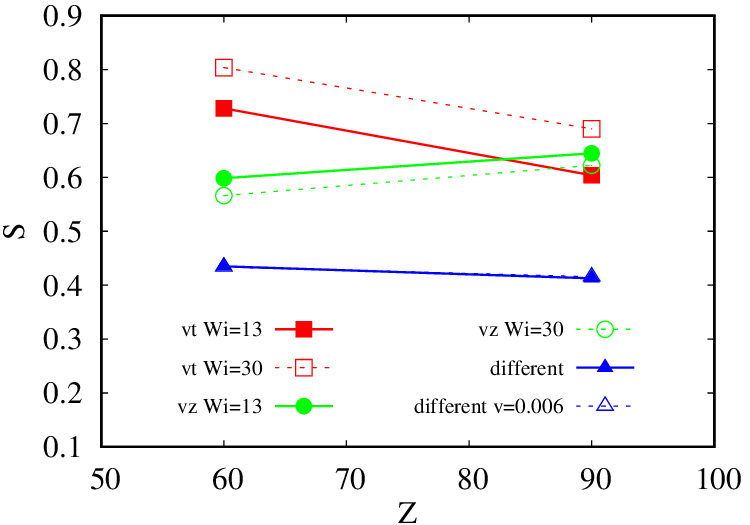}	
		\caption{Similarity calculations for all the systems we compared. "vt" corresponds to systems where $v$ and $\tau$ are inversely proportional. "vz" corresponds to systems where $v$ and $Z$ are proportional. "different" corresponds to systems where $Wi$ are different.}
		\label{fig:result}
	\end{figure*}

	Through quantitative calculation of similarity, we obtained the results shown in Fig.~\ref{fig:result}. From Fig.~\ref{fig:result}, we can see that the system similarity ranges from high to low as follows: $v$ and $\tau$ change inversely, $v$ and $Z$ change positively, and different $Wi$. When $v$ and $Z$ change positively, the similarity is lower than when $v$ and $\tau$ change inversely. There could be several reasons for this: 1. The dimensionless quantity $R/Z$ is an observable quantity ($1/40$) in the simulation, but we do not change the particle size $R$ proportionally with $Z$ because such a change would make the system equivalent in the simulation. Therefore, the change in $R/Z$ reduces the similarity of the system. Specifically, we can observe that the similarity is lower when the $Z$ changes from $40 \sigma$ to $60 \sigma$, but becomes higher when the $Z$ changes from $60 \sigma$ to $90 \sigma$. This is because, for a system with a larger $Z$, the value of $R/Z$ tends to be smaller, resulting in a smaller effect on the system and thus making the systems more similar. It is worth noting that this value is very small and can be ignored in macro systems. 2. Due to $\xi$ changes in different systems, it is necessary to keep $\tau$ unchanged by controlling $N$. However, since $N$ is discrete and different values of $N$ have a greate impact on the relaxation time, it is impossible to accurately control the relaxation time, resulting in changes in $Wi$. Conversely, when $v$ and $\tau$ vary inversely, since the plate thickness $Z$ is the same for the different systems, the dimensionless quantity $R/Z$, although it is an observable quantity, has a constant value. Therefore, it does not reduce the degree of similarity of the system, and the similarity does not increase due to the increase in size. Additionally, since $N$ does not change, $Wi$ of different systems are closer to each other. In addition, when $v$ and $\tau$ are varied inversely, the fluctuations in the similarity of the different systems are due to the stochastic nature of the fluid motion.

	\section{Summary and remarks}
	
	To study the mechanism of flow marks in injection-molded products and eliminate them, we would like to be able to visualize the flow process of the melt responsible for the formation of flow marks. However, this process is too macroscopic for molecular dynamics simulations and the computational effort cannot be matched. Therefore, we propose scaling down the flow field in equal proportions and then visualizing this process using molecular dynamics simulations.
	
	Using dimensional analysis, we find the dimensionless quantity that governs the geometry of this flow field. We conclude that when the mold geometry is proportionally varied, with fixed melt and mold materials and fixed initial temperatures of the melt and mold, the flow field geometry will be solely controlled by the Wiesenberg number $Wi$. Keeping $Wi$ constant, modifying the injection speed, adjusting the relaxation time of the polypropylene melt, or scaling the mold will yield similar geometric shapes of the flow field.
		
	Theoretically, based on the above conclusions, we can scale the flow field proportionally and visualize the melt flow process that leads to the formation of flow marks using molecular dynamics simulations. However, before doing so, we need to verify the correctness of the above conclusions. We verified the above conclusions through microscopic molecular dynamics simulations with the same values of dimensionless quantities as the experimental injection molding process (Some of these values differ from the experiment, such as $\frac{w}{Z}, \frac{L}{Z}, \frac{\xi^{\prime}}{\xi}$, etc. However, they are of the same order of magnitude as the experiment and thus have the same dimensional analysis).
	
	Before verifying whether the flow field is controlled by $Wi$ through simulations, we made some preparations. First, we consider that for the simulated system, we also need to match the dimensionless quantity $k T/v^2 \xi t$ of the flow in the pipe so that the flow rate at the outlet of the pipeline can meet the scaling (for the experiments, this quantity is very small, and the velocity driving the melt scales equally, the outlet flow rate automatically satisfies the scaling). Due to the existence of dimensionless numbers in the pipe, our only strategy for the choice of scaling relationships for the simulation parameters is as follows: 1. When $v$ and $Z$ increase proportionally, reduce $\xi$ by the square multiple of the increase in $v$ (increasing $T$ is equivalent to decreasing $\xi$), changing $N$ ensures that $\tau$ remains constant, and the program running time $t$ remains constant. 2. When $v$ and $\tau$ are varied inversely, $Z$ remains unchanged, and if $v$ decreases, then $\xi$ increases in the same proportion. Therefore, $\tau$ increases in the same proportion, and the program running time $t$ increases in the same proportion. To determine the value of $N$ when $v$ and $Z$ increase proportionally, the relationship between the degree of polymerization and the relaxation time is obtained by measuring the mean-square displacement of the unit. We match the other adjustable parameters with the experiment to make each dimensionless quantity the same as the experimental value or be a very large or very small quantity as in the macroscopic experiment.
	
	The simulation results show that the flow fields are similar for systems with the same $Wi$, and significantly different for systems with different $Wi$. Through quantitative calculation of similarity, we can see that the system similarity ranges from high to low as follows: $v$ and $\tau$ change inversely, $v$ and $Z$ change positively, and different $Wi$. When $v$ and $Z$ change positively, the similarity is lower than when $v$ and $\tau$ change inversely. This may be due to the variation of the dimensionless quantity $R/Z$ and the error caused by changing $N$. We can see that the similarity is lower when $Z$ changes from $40 \sigma$ to $60 \sigma$, but becomes higher when $Z$ changes from $60 \sigma$ to $90 \sigma$. This shows that the effect of $R/Z$ on the simulated system diminishes as this dimensionless number tends to be small. The results of the simulation validate the conclusion of the dimensional analysis to some extent. Furthermore, we will improve the system to enhance its similarity, and select the same parameters as the experimental injection molding process for molecular dynamics simulation to visualize the flow mark formation process. Since the formation of flow marks has an induction length,\cite{maeda2007flow} a longer simulation time will be required. After the system is completed, it will be possible to visualize the process of flow mark formation, which will help to advance the analytical theory and eliminate the flow marks in production and experiments. This work also illustrates that the methodology of quantitative analysis plus simulation may be applied to a wider range of other systems, scaling down large systems and thus significantly reducing their computational effort.

	\section{Acknowledgements}
	
	Financial support by the National Natural Science Foundation of China (grant nos. xxxxxx and xxxxxx) is gratefully acknowledged.

	\section{Author Declarations}
	
	\textit{Conflict of Interest}
	
	The authors have no conflicts of interest to disclose.

	\section{Data Availability}
	
	The data that support the findings of this study are available within the article.
	
	\bibliography{dimensional_analysis}

\end{document}